\documentclass[fleqn,usenatbib]{mnras}

\usepackage{newtxtext,newtxmath}

\usepackage[T1]{fontenc}

\DeclareRobustCommand{\VAN}[3]{#2}
\let\VANthebibliography\thebibliography
\def\thebibliography{\DeclareRobustCommand{\VAN}[3]{##3}\VANthebibliography}

\usepackage{graphicx}   
\usepackage{amsmath}	
\usepackage{amssymb}	
\usepackage{pdflscape}

\title[Metallicity distribution of the GSS progenitor]{Metallicity distribution of the progenitor of the
Giant Stellar Stream\\ in the Andromeda Galaxy}

\author[S. Milo{\v s}evi{\' c} et al.]{
S. Milo{\v s}evi{\' c},$^{1}$\thanks{E-mail: stanislav@matf.bg.ac.rs}
M. Mi{\' c}i{\' c},$^{2}$ and
G. F. Lewis$^{3}$
\\
$^{1}$University of Belgrade, Department of Astronomy at Faculty of Mathematics, Belgrade, Serbia \\
$^{2}$Astronomical Observatory of Belgrade, Belgrade, Serbia\\
$^{3}$Sydney Institute for Astronomy, School of Physics, A28, The University of Sydney, NSW 2006, Australia
}

\date{Accepted XXX. Received YYY; in original form ZZZ}

\pubyear{2022}

\begin{document}
\label{firstpage}
\pagerange{\pageref{firstpage}--\pageref{lastpage}}
\maketitle

\begin{abstract}
The Giant Stellar Stream (GSS) is a prominent tidal feature in the halo of the Andromeda Galaxy (M31), representing the ongoing destruction of a satellite galaxy. In this paper, we investigate the formation of the GSS through detailed numerical simulations of the tidal disruption of a progenitor system.
Assuming that the stream was created in a single merger event between M31 and a dwarf spheroidal galaxy with stellar mass of $10^{9}M_{\odot}$, we successfully reproduce the dynamical properties of the GSS.
As the metallicity distribution along the stream has been well determined from the observations (PAndAS and AMIGA data sets), we use Monte Carlo simulations to reconstruct the original metallicity distribution of the dwarf progenitor. We find that a progenitor dwarf galaxy with a negative radial metallicity gradient, $\Delta$FeH = -$0.3 \pm 0.2$, reproduces the observed GSS properties at a time between 2.4 and 2.9 Gyrs into the merger. We also show that the observed double peak metallicity distribution along the stream is a transitory structure caused by unique merger circumstances where two groups of streaming stars are moving in opposite directions, intersecting to produce the peaks.  
\end{abstract}





\section{Introduction}
Spiral galaxies can exhibit rich structures, which are formed in merger events, flybys, or other dynamical and evolutionary processes, such as stellar streams, shells, warps, rings, tidal tales. Flybys and mergers can also produce instabilities in disks of spiral galaxies. The properties of these various structures depend upon many parameters, in particular the properties of galaxies in merger event: mass ratio, orbital circularity, angular momentum vector at the time of accretion onto the host halo (different orientation of disk of the host galaxy), the morphology of each galaxy (\citealp{Hernquist1988, Hernquist1989, Lang}). 

M31 galaxy is the largest member of our Local Group of galaxies, being a large spiral similar to our own Milky Way.
%
 Due to its proximity, M31 has been studied in depth, providing an unparalleled view of a large spiral. In particular, over the last two decades, the Pan-Andromeda Archaeological Survey (PAndAS) has provided a panoramic view of the halo of Andromeda, revealing a wealth of substructures. (\citealp{Ibata2007, Ibata2014, Martin, McConnachie2018}). 
 Amongst the most prominent structure is the Giant Stellar (or Southern) Stream. Originally identified by \citet{Ibata2001}, numerous observations of the GSS have been undertaken (\citealp{Ferguson, Bellazzini, McConnachie, Conn}), mapping the spatial distribution and heliocentric distance to the stream stars, with (\citealp{Ibata2004, Guhathakurta, Gilbert2009, Gilbert2018}) determining the velocity distribution along the stream. The kinematical picture of the GSS is very complex. A second kinematically cold component was discovered in the line-of-sight velocity distribution of the GSS (\citealp{Kalirai2006a, Gilbert2009}) that is separated from the primary GSS component by 100 km/s over a radial range of $\sim$ 7 kpc, and may represent a bifurcation in the line-of-sight velocities of stars in the GSS. The GSS extends outward of the central region of M31 for 6 degrees on the sky corresponding to a projected radius of 80 kpc (\citealp{Cohen}). The luminosity of the stream is $3.4 \times 10^{7} L_{\odot}$ and stellar mass $M_{GSS}\approx 2.4 \times 10^{8}M_{\odot}$ (\citealp{Ibata2001, Fardal2006}). 
 
 Chemical abundances was investigated both, photometric (\citealp{Ibata2007, Brown2006a, Brown2009}) and spectroscopic (\citealp{Koch, Gilbert2009, Gilbert2014}). Recent photometric results (for spectroscopically confirmed M31 RGB stars) in chemical abundances and kinematics are given in SPLASH (\citealp{Guhathakurta, Kalirai2006a, Gilbert2009, Gilbert2014}) and, spectroscopic analyses in the Elemental Abundances in M31 collaborations (\citealp{Gilbert2019, Escala2020, Escala2021}). The work of \citet{Ibata2007} shows a difference in metallicity between the core and envelope of the GSS. This was confirmed spectroscopically in \citet{Gilbert2009} revealing a decrease in metallicity from core to the envelope. In \citet{Conn} and \citet{Cohen} are presented metallicity values along the GSS. From 19 observational fields, it is shown two gradients across the GSS. From the inner part most close to the center of M31, metallicity takes a value of -0.7 and increases to -0.2 in the central region and then decreases in the outer part reaching the value of -1. In the central part of the GSS double-peaked metallicity distribution is observed (\citealp{Conn}) and the nature of this is still unclear.   
 
 To explore the formation scenario in detail a significant number of simulation realizations have been undertaken. Minor merger scenarios were investigated in the works of \citet{Fardal2006, Fardal2007}, proposing that the progenitor of the GSS was a satellite galaxy on highly radial orbit with stellar mass $M \sim 10^{9} M_{\odot}$, without dark matter and finding good agreement between simulated and observed heliocentric distances and velocity distribution in the stream and also the orientation of the GSS. It is considered that satellite was already tidally stripped and the timescale of the merger was several hundred Myr. Other structures like the NE and W shelves, discovered by \citet{Ferguson}, were reproduced, and also the position of the remnant of the satellite was in the vicinity of the NE shelf (\citealp{Fardal2013}). In \citet{Mori} the Stream and shelves structures were also reproduced in N-body simulation, and the nature of the satellite was discussed.

 A single-merger scenario was proposed in the work of \citet{Sadoun}, where M31 was represented as a live N-body system, and the turnaround radius of the satellite galaxy was 200 kpc with a null initial velocity. In \citet{Sadoun}, a model for satellite galaxy was introduced with a 20 times more massive dark matter halo than the baryonic matter content. In this work the best time for the GSS formation was 2.7 Gyr. The position of the core of the progenitor was in the region of the NE shelf.   
Furthermore, the N-body representation of M31 accounts for dynamical friction as an important dynamical phenomenon for galaxies, which is only approximated in static representations. The single merger scenario was considered in \citet{DSouza2018}, but with a more massive satellite galaxy with stellar mass $M=2,5 \times 10^{10}M_{\odot}$ and timescale of the merger $\sim$ 2 Gyr. Different morphologies were suggested for the progenitor galaxy. In \citet{Fardal2008} satellite with and without disc was investigated, while \citet{Sadoun} found that satellite without disc better reproduces observed depth of GSS and velocity distribution. In \citet{Miki} different types of progenitor were proposed: dwarf elliptical, dwarf irregular, and small spiral, while in \citet{Kirihara2017} thick disc progenitor was analyzed. The major merger scenario was introduced by \citet{Hammer2010, Hammer2013} with analyses of single and multiple mergers of M31 and its satellites. In \citet{Hammer2018} major merger scenario was also represented with hydrodynamical simulation and the mergers occurred over 2-3 Gyr. Morphology of the progenitor is still an open question.

 The origin of the metallicity gradient in the GSS is connected with the properties of the progenitor. Gradients in metallicity have been determined in spiral galaxies and dwarf galaxies (\citealp{Koleva2009a, Koleva2009b, Spolaor, Ross}). Also, the metallicity content of a galaxy depends on the stellar mass, providing an additional constraint for the mass through the metallicity mass relation (\citealp{Gallazzi, Panter, Gonzalez, Kirby}).
 
 The connection between observed metallicity in the GSS and its possible progenitor is presented in \citet{Fardal2008}. In the work of \citet{Miki} motivated with observations of metallicity gradient in dwarf elliptical (\citealp{Koleva2009a, Koleva2009b, Spolaor}) they investigated both: positive and negative metallicity gradient in progenitor. The results of the simulation were compared with several observed fields in the radial direction (\citealp{Guhathakurta, Kalirai2006a, Kalirai2006b, Koch, Gilbert2009}). A small number of fields were located between $R_{proj}$ of 20 and 40 kpc. Radial metallicity distribution observed in the GSS after the merger can give information about metallicity distribution in progenitor (\citealp{Miki, Panithanpaisal}). After photometric observations of \citet{Conn} and \citet{Cohen} it became possible to compare simulated metallicity distribution with observed one across the radial extent of the GSS up the 6 degrees, in the 19 observational fields. 
 
 In this work, we undertake N-body simulations of the merger of M31 with a dwarf galaxy progenitor, and we reproduce the formation of the GSS in a similar way as \citet{Sadoun}. The baryonic mass of the progenitor galaxy is the same as in \citet{Fardal2007} and it is constrained by the mass of the GSS, but there is uncertainty in determining its dark halo mass. After we successfully reproduce dynamical properties of the GSS, we use Monte Carlo (MC) simulations to find the initial metallicity gradient in the dwarf progenitor, which leads to the observed metallicity distribution along the stream, after the merger. Here, we attempt to explain metallicity distribution along the GSS that is observed in works of \citet{Conn} and \citet{Cohen} by probing different metallicity distributions in the progenitor galaxy.  As dwarf galaxies generally contain very little gas due to the action of galactic winds, tidal stripping, and dynamical friction, gas is neglected in this work. 
 
 This paper is organized as follows: In Section 2 we introduce method and N-body models for M31 and the satellite progenitor of the GSS. Also, we present the MC method for finding the initial metallicity distribution in the dwarf progenitor galaxy. In Section 3 we present the main results based on the comparison of simulations to observations. In Section 4 we discuss our results and conclude.

 \begin{figure*}
\centering
\includegraphics[scale=0.9]{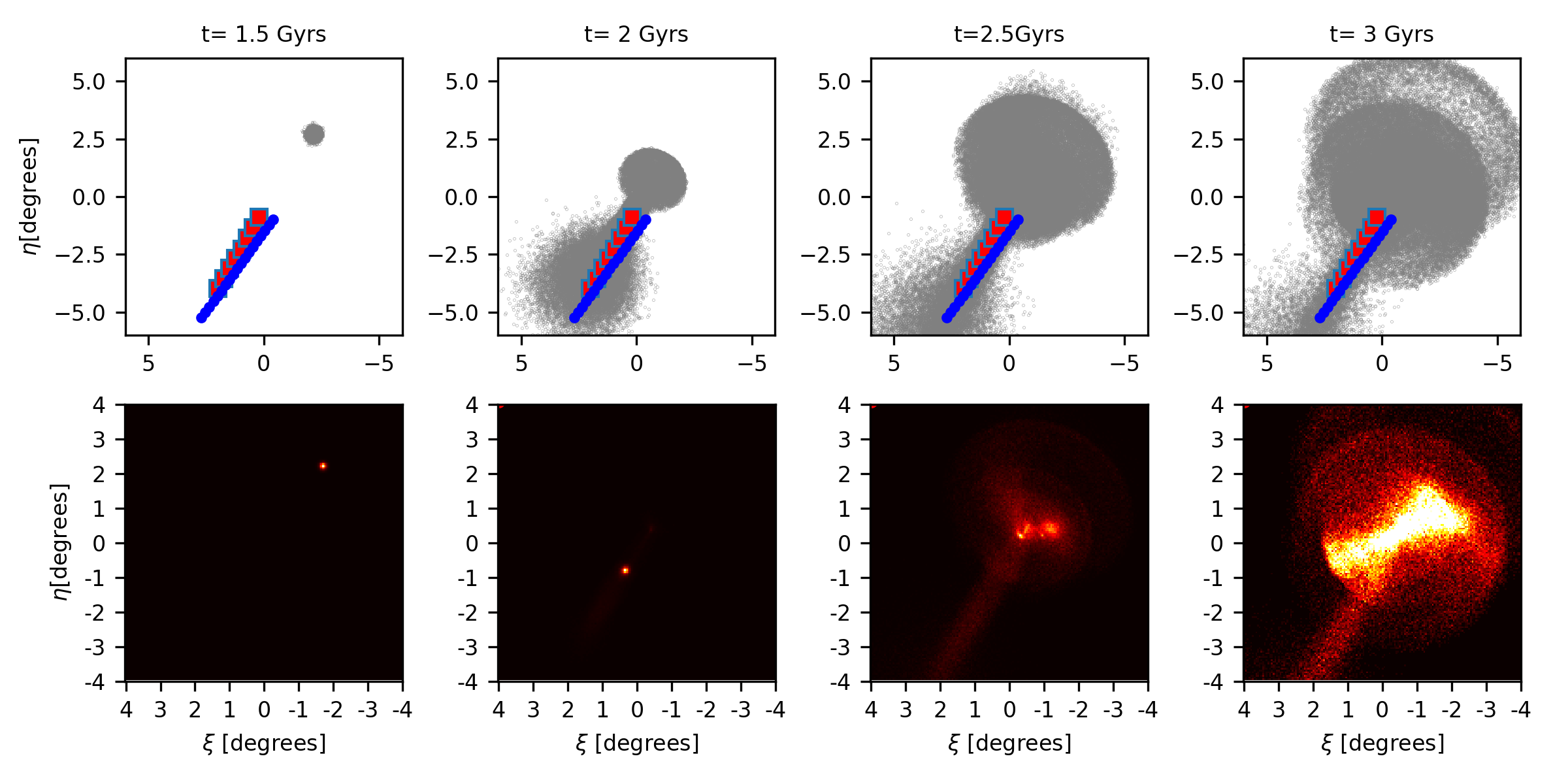}
\caption{The dwarf progenitor galaxy presented projected on the sky for four epochs after the merger begins, at 1.5, 2.0, 2.5, and 3.0 Gyrs. The top panels show all of the dwarf progenitor particles in the  simulation. Red boxes are observed fields in \citet{McConnachie} and blue cirles are fields from \citet{Conn}. The bottom panels show the mass surface density of the same galaxy. M31 particles are removed from all panels so that stellar structure formation can be clearly seen.}
\end{figure*}

\section{Methods}
 
\subsection{N-body model for the Andromeda Galaxy}
 M31 is a spiral galaxy and, we assume three key components: bulge, disk, and dark matter halo. We use the N-body model for M31 similar to that one used in previous works (\citealp{Geehan, Sadoun}). For generating a pure N-body model we use GalactICS package \citep{Widrow} that computes gravitational potential for a given mass model and distribution function for each component of the galaxy. We assume a total mass of M31 to be $\sim 10^{12} M_{\odot}$ and we represent it by 467081 particles. With this total number of particles we have the same mass resolution as in \citet{Sadoun}: $10^{5} M_{\odot}$ for baryonic particles and for $10^{6} M_{\odot}$ dark matter particles. The particle mass in each component and number of particles are given in Table 1.\\
\indent The bulge is represented with the Prugniel-Simien profile \citep{Widrow}, which is a de-projected Sersic profile. This profile has 
 $r^{1/n}$ law.:
\begin{equation}
\rho_{b}=\rho_{b0} \left( \frac{r}{r_{b}} \right) \textrm{exp} \left( r/r_{b} \right)^{-1/n}.
\end{equation}
Here, $\rho_{b0}$ is density at $r=r_{b}$, and $r_{b}$ is a scale radius for the bulge. The value for $n$ is 1.8.   

The disk is represented by a combination of two profiles: exponential profile of surface density in the x-y plane and $\rm sech^{2}$ law in the vertical, z-direction.  
The exponential profile is given with (\citealp{Geehan, Sadoun}):
\begin{equation}
\Sigma(R)=\frac{M_{d}}{2\pi R_{d}^{2}} e^{-\frac{R}{R_{d}}}
\end{equation}
\noindent Here, $M_{d}$ is the total mass of the disk, $\Sigma$ is surface density and $R_{d}$ is disk scale radius. In the last two equations, $r$ is the spherical radius, and $R$ is the cylindrical radius.

The $\rm sech^{2}$ profile is used in vertical (z) direction \citep{Sadoun} and combined profile is given by:
\begin{equation}
\rho(R,z)=\frac{\Sigma(R)}{2z_{0}} \textrm{sech}^{2} \left( \frac{z}{z_{0}} \right)
\end{equation}
\noindent Here, $z_{0}$ is the scale height of the disk. The inclination of the disk is $77^{o}$ and position angle is $37^{o}$ \citep{Fardal2007}, and the heliocentric distance to Andromeda is taken to be 785 kpc. 

A spherical dark matter halo is represented with Navarro-Frenk-White profile \citep{Navarro}:
\begin{equation}
\rho_{h}(r)=\frac{\rho_{0}}{\frac{r}{r_{h}}\left(1+\frac{r}{r_{h}}\right)^{2}}
\end{equation}
\noindent but more general form is given in GalactICS with an additional term for truncation at some point.
\begin{equation}
\rho(r)=\frac{2^{2-\alpha}\sigma_{h}^{2}}{4\pi r_{s}^{2}}\frac{\rho_{0}}{(r/r_{s})^{\alpha}(1+r/r_{s})^{3-\alpha}}\frac{1}{2} \textrm{erfc} \bigg{(}\frac{r-r_{h}}{\sqrt{2} \delta_{r_{h}}}\bigg{)}
\end{equation}
\noindent Here, $r_{h}$ is radius of halo and the value at which density starts to decrease, $\sigma_{h}$ is typical velocity, $\delta_{r_{h}}$ is distance along which density falls to zero, $r_{s}$ is scale radius for halo and $\alpha$ is exponent in NFW profile and we took $\alpha = 1$. The tidal radius $r_{200}$ is taken to be approximately equal to the distance where density drops 200 times in comparison to central density. The total mass of the halo inside $r_{200}$ radius is $M_{200}=8.8\times 10^{11}M_{\odot}$. The values of parameters given in Table 1, are used in GalactICS in the way described in \citet{Widrow}, where N-body models of M31 are presented with parameter values motivated from observational works of \citet{Kent}, \citet{Braun} and \citet{Widrow2003}.
Adopted parameters are similar as in \citet{Sadoun}.

\begin{table*}
\centering 
\begin{center}
\begin{tabular}{|c|c|c|c|c|c|c|c|} 
\hline
\hline
component & m [$M_{\odot}$] & N \\
\hline
Bulge &  $3.36 \times 10^{5}$  & 96247 & $r_{b}$ = 1,23 kpc & $\sigma_{b}=$ 393 km/s & $M_{b}= 3.2 \times 10^{10} M_{\odot}$   \\
\hline
Disk &  $3.36 \times 10^{5}$ & 108929 & $R_{d}$ = 6.82 & $z_{0}$= 0.57 & $M_{d}=$ $3.66 \times 10^{10} M_{\odot}$\\
\hline
Halo & $3.36 \times 10^{6}$ & 261905 & $r_{h}$= 122.5 kpc & $r_{s}=$ 8 kpc & $M_{h} = 8.8 \times 10^{11} M_{\odot}$ & $\delta_{r_{h}}$= 12 kpc & $\sigma_{h}$ = 525 km/s  \\
\hline
\end{tabular}
\end{center}
\caption{The values of the parameters for the N-body model of M31 used in GalactICS, where m is mass of one particle, and N is the number of particles in each component. Similar values are used in \citet{Geehan}, \citet{Fardal2007} and \citet{Sadoun}.}
\end{table*}

\subsection{N-body model for progenitor galaxy}
There have been many different models for satellite galaxy as the GSS progenitor, with different morphologies investigated to explain the formation of the GSS and other stellar structures. In works of \citealp{Fardal2007, Fardal2008} and \citet{Sadoun} a progenitor with disk and spherical progenitor were analyzed. As it is suggested in \citet{Sadoun}, we used a spherical model for a GSS progenitor, which includes a spherical bulge and dark matter halo. 

We model the bulge of the progenitor galaxy as represented in Equation 1, but with a mass of $2.18 \times 10^{9} M_{\odot}$. This mass is the mass of the baryonic matter in the satellite galaxy and it is the same as in \citet{Fardal2006} and \citet{Sadoun}. The baryonic part of the progenitor is well constrained by the mass of the GSS which is $2.4 \times 10^{8} M_{\odot}$ (\citealp{Ibata2001, Fardal2006}), so after the merger event, particles in the region of GSS give the mass suggested in observations. Unlike the baryonic part, the mass of the halo is not well determined. We assume a cosmologically relevant merger scenario and stellar mass/dark matter mass ratio as it is represented for the Local group galaxies in \citet{Brook}. The dark matter halo is 20 times more massive than the baryonic part and the mass of the halo is $4.13 \times 10^{10} M_{\odot}$. The density profile for halo is NFW with the truncation parameter as described in equation (5). The number of the particles for each component and their masses are given in Table 2. These values are similar to one given in \citet{Sadoun}.


\begin{table*}
        \label{tab:landscape}
        \begin{tabular}{lccccr} 
            \hline
            \hline
            component & m [$M_{\odot}$] & N  \\
            \hline
            Baryonic matter & $1.66 \times 10^{4}$ & 131072 & $r_{s}=$ 1.03 kpc & $\sigma_{s}=$ 93 km/s & $M_{s}=2.18 \times 10^{9} M_{\odot}$ \\
            \hline
            Dark matter halo & $1.66 \times 10^{5}$ & 248809 & $r_{h}=$ 5 kpc & $\sigma_{h}=$ 242 km/s & $M_{h}=4.13 \times 10^{10}M_{\odot}$\\
             \hline
        \end{tabular}
        \caption{The values of the parameters for the N-body model of progenitor galaxy, where m is mass of one particle, and N is the number of particles in each component. Similar values for the baryonic matter are used in \citet{Sadoun}.}
\end{table*}

\subsection{Parameters of the merger event}
 We use the N-body representation of M31 and for the progenitor of the GSS but without a disk component for the latter. In \citet{Fardal2007} satellite galaxy without dark matter halo, that was earlier disrupted due to tidally striping, was analyzed and observational quantities were reproduced. In \citet{Sadoun} a progenitor with a dark matter, on its turnaround radius of 200kpc started to fall on M31. Orbital parameters are dependent on the mass of the progenitor and its velocity. They can vary with changing mass, which is described in \citet{Hammer2018}, where different mass ratios were assumed for the different merger scenarios for the formation of the GSS, whether minor or major merger scenario is investigated. As our model contains a similar amount of dark matter as the model represented in \citet{Sadoun}, and similar total mass, initial distance in merger event is also similar. The initial coordinates are:\\    
\begin{center}
(x,y,z) = (-84.41,108.47,-156.08) kpc
\end{center}

We ran our simulations with Gadget2 dynamical code \citep{Springel}, with the numerical results from simulations compared with observations. Gadget2 is a cosmological simulation code that can apply the TreePM algorithm and computes short-range forces with the "tree" method and longe range forces with Fourier methods \citep{Springel}. The orientation of the stream was compared to the observed properties given in \citet{McConnachie}, and heliocentric distances with those given in \citet{McConnachie} and \citet{Conn}. Velocities along the stream were compared to the measurements (\citealp{Ibata2004, Guhathakurta, Gilbert2009, Gilbert2018}). We assumed that the merger event lasts between 2 and 3 Gyrs (\citealp{Sadoun,Hammer2018}). 

\subsection{Initial metallicity distribution of the dwarf progenitor galaxy}
To explain the metallicity distribution in the GSS described in \citet{Conn} and \citet{Cohen}, we assume that the initial distribution in the progenitor before the merger mainly impacts the distribution along the GSS after it was formed in merger event. There are two observed gradients along the GSS: a metallicity increases from its minimum observed value of - 0.7 (when the GSS is closest to M31) to the maximum value of -0.2 (3 degrees distance from the center of M31), then decreases toward the end of the GSS where it has the value of -0.8 (\citealp{Conn, Cohen}). These first observations of the metallicity along the stream are derived from color-magnitude diagrams. Metallicity values in fields far away from the inner part of the Stream, are more affected by foreground contamination \citep{Cohen} and these values could differ according to different cleaning techniques. As earlier mentioned, before these observations, a metallicity gradient in the direction perpendicular to the GSS was known, with higher metallicity in the core region and lower in envelope (\citealp{Ibata2007, Gilbert2009}). 

We searched for the required initial metallicity distribution in the progenitor necessary to reproduce the observed metallicity in the GSS after the merger, along with the two gradients. We first assume that the initial metallicity distribution is a linearly decreasing function from the center of the progenitor. The maximum and minimum values of metallicity in progenitor are constrained by the observed values in the GSS. Also, distribution depends on the morphology of the dwarf galaxy. We assume the dSph model for the GSS progenitor. Each particle in the progenitor can be tagged (assigned) with a specific value of metallicity from the initial distribution. Particles carry their metallicities through the simulation run. After the simulation is finished, and the GSS is formed, we extract particles that make the GSS and read their metallicities. Based on that, we construct metallicity distribution in simulated GSS and compare it to the observed one. Also, we calculate mean metallicity in the observation fields using the coordinates given in \citet{Conn} and compare these values to the observed values in these fields. We test the final metallicity distributions along the stream for the several post-merger moments (between 2 and 3 Gyrs). Then we vary the initial metallicity distribution in the progenitor and use Monte Carlo (MC) simulation to find which one produces the match between the simulated and the observed metallicity distribution in the GSS. The scheme of MC simulation includes:
\begin{enumerate}
\item Linearly decreasing function for the initial metallicity distribution in the progenitor 
\item Progenitor is divided in to shells 
\item In each shell mean metallicity [Fe/H]$_{mean}$ is taken from the linear function in (i) 
\item We assume a Gaussian distribution of metallicity in each shell with mean value [Fe/H]$_{mean}$ 
\item For each particle in one shell, we randomly draw one metallicity value from the Gaussian distribution
\item Repeat the procedure 1000 times 
\item Track the particles from the progenitor to the GSS and read their metallicities. Match particles' positions to the fields in the observations. Calculate mean metallicities and error bars for each observation field 
\item Compare calculated metallicities to the observed ones. 
\end{enumerate}
We use a metallicity of -0.2 at the center of the dwarf galaxy and then we linearly decrease it to -1.5 in the outer part of the galaxy as described in the MC procedure. We also changed these values and tested metallicity in the center from 0 to -0.3 and in the outer part from -1 to -1.8. These values are motivated from observations given in \citet{Conn}. For each shell we took $\sigma = 0.4$ for gaussian distribution of metallicity in the shell, and that value is double of the values presented in \citet{Conn}. The final distribution in these tests is well covered since for the different initial maximum and minimum values in the center and the outer part of the GSS progenitor, we reproduced lower and higher values for metallicity compared to the observed one. Mass, the morphology of the progenitor, together with the circumstances of the merger event, are the fixed parameters in our attempt to model GSS metallicity distribution. The only variable parameter is the initial metallicity distribution. 

\begin{figure*}
\centering
\includegraphics[scale=0.99]{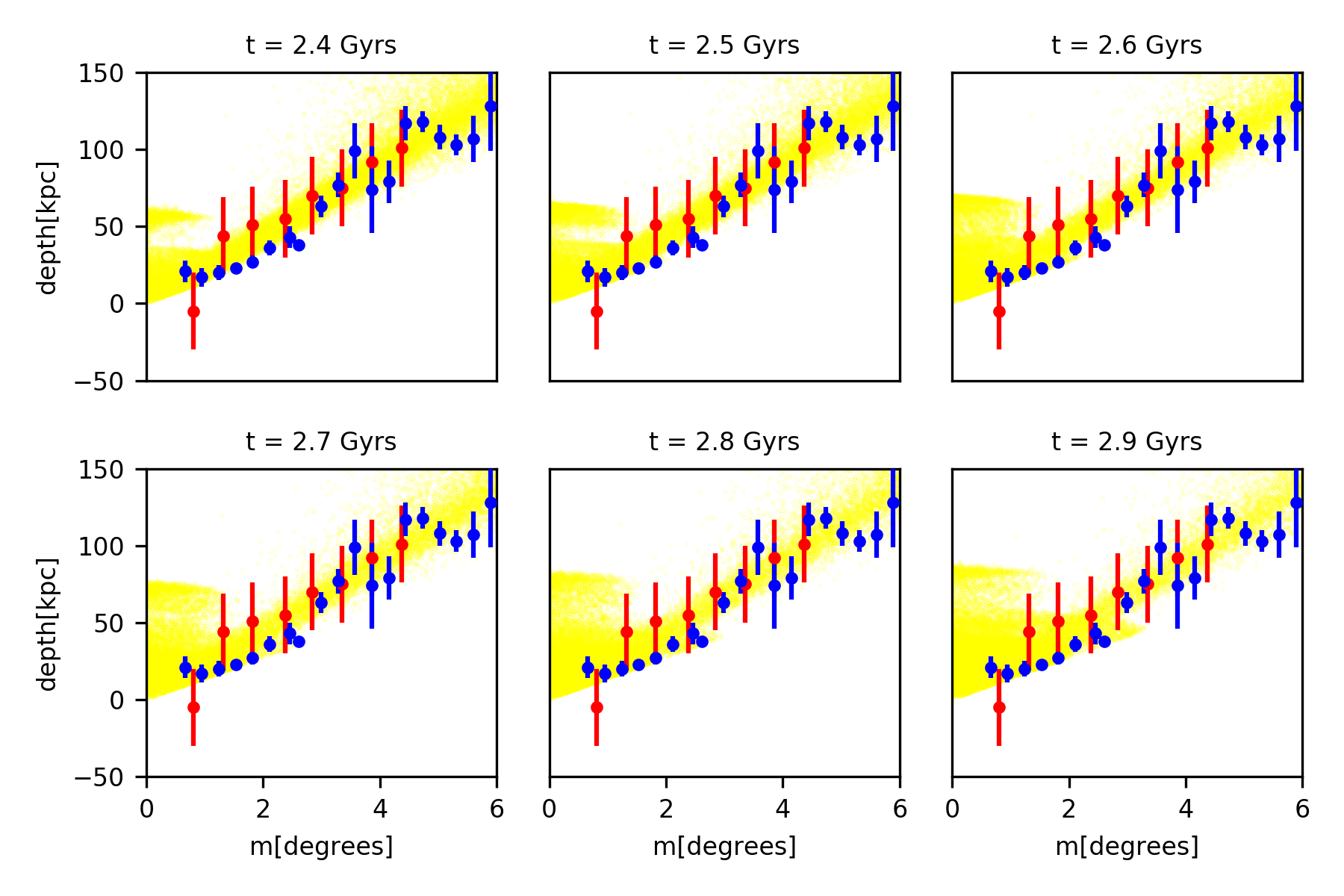}
\caption{Distances from the center of M31 on y-axis. Bars are observed values for stars from \citet{McConnachie} (red) and \citet{Conn} (blue). Dots are simulated particle distances in our N-body simulation. Panels show different moments into the merger, from 2.4 to 2.9 Gyrs.}
\end{figure*}

\section{Results}
We present two groups of results, with the first focusing upon our numerical simulation where we show reproduced dynamical properties of the GSS, its position in the sky, depth and velocities along the GSS. The second explores our MC simulation which is painted on top of our numerical simulation. These results are related to the progenitor's initial metallicity distribution and the GSS's final metallicity distribution.

\subsection{Results of the numerical simulation}
We assume pure N-body models for both galaxies, M31 and dwarf spheroidal. The dwarf spheroidal galaxy starts its trajectory with null velocity from the initial position which is 207.97 kpc from the center of M31 \citep[e.g.][]{Sadoun}. As the dwarf approaches the central region of the host galaxy, the dark matter halo starts to ``peel off'' because of the tidal interactions with the halo of M31. The baryonic core of dwarf survives and passes through the central region of the potential well after 1.5 Gyrs. Due to dynamical friction and tidal processes, it starts to slow down and leave baryonic material in the length of 5 degrees in the sky, forming the GSS. Because of the gravitational attraction of M31, the smaller core of the dwarf galaxy turns back and it ends in the vicinity of the NE shelf.  

\begin{figure}
\centering
\includegraphics[scale=0.55]{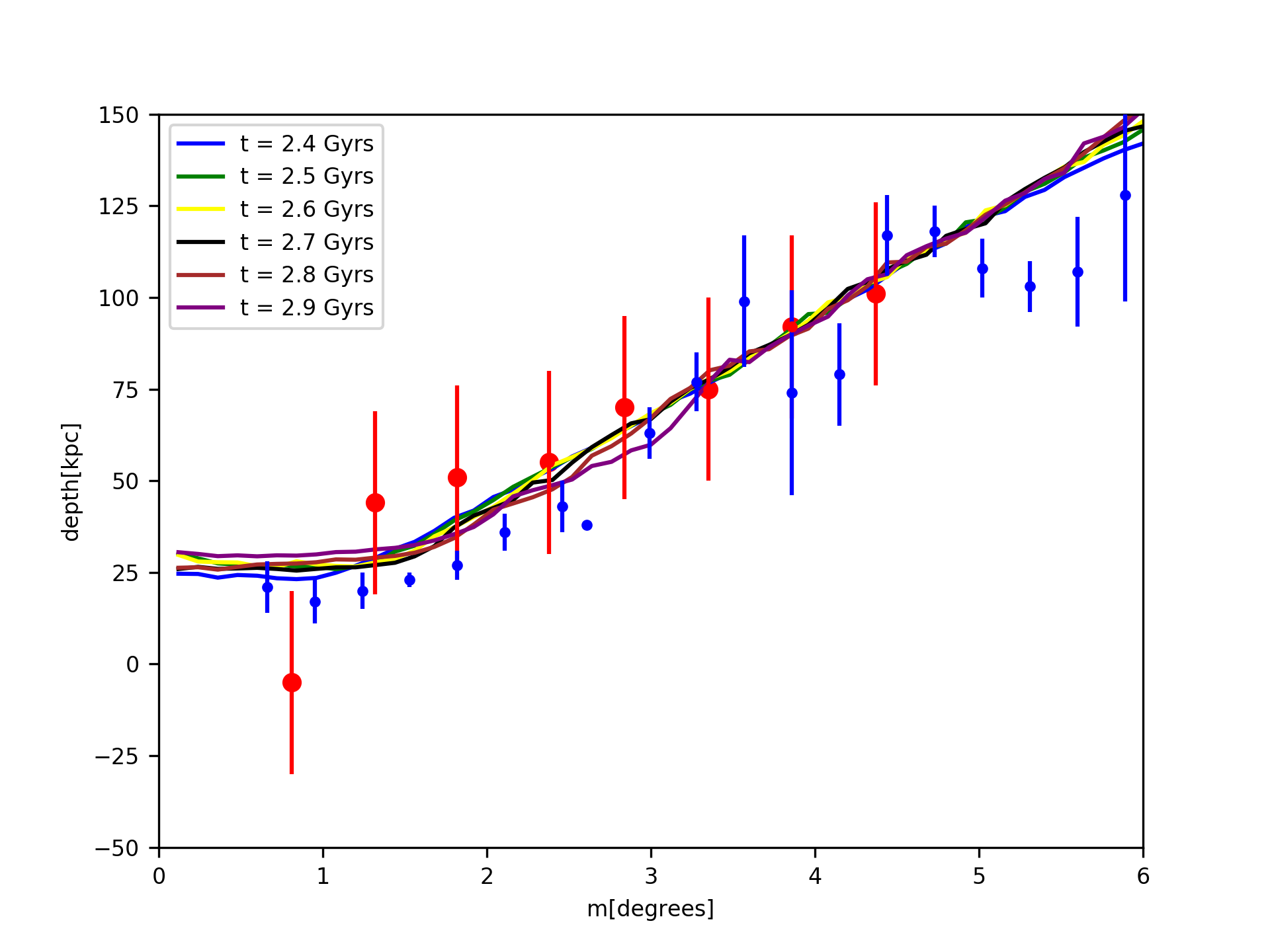}
\caption{Mean depth of stars in eight observational fields presented with error bars (red from \citet{McConnachie} and blue from \citet{Conn}, and mean depth for particles in the simulation presented with lines. Each line represents a snapshot during the last 500 Myrs of the GSS formation.}
\end{figure}

\subsubsection{Orientation of the GSS}
In Figure 1 the merger history is represented from the beginning until 3 Gyrs into the merger. The dwarf progenitor's particles are represented in the upper row of panels, and surface density maps in the lower row, for the same moments in time. In this figure we can see the dwarf progenitor approaching the centre of M31 on almost radial trajectory at 1.5 Gyrs (first column panels). At 2 Gyrs into the merger (second column panels), dwarf progenitor has passed through the central region of M31, shredded its stellar material toward what is later going to be the GSS, and its on a way back toward a second pericentric passage. At the end, the core remnant of the dwarf progenitor ends in the vicinity of the NE shelf of M31. 
M31 particles are not presented in the figure for the better visibility. The GSS clearly stands out as a stellar structure which forms in the southern region of Andromeda. We compare the orientation of the GSS to the observations given in \citet{McConnachie}. Eight observed fields for comparison to our simulated data can be seen in the left panel. These fields are at a different radial distances along the stream. The orientation of the simulated GSS in the sky is constrained by the observed fields.
The morphology of the GSS is represented in Figure 1 as we can see it in the sky projected on x-y plane, with x-axis oriented from right to left, and z direction oriented along the line of sight. These coordinates are translated in angular coordinates $\xi$ and $\eta$ according to equations \citep{Fardal2007}:
\begin{equation}
\xi = x/(z_{g} + z) \\
\eta = y/(z_{g} + z)
\end{equation}
where $z_{g}$ is the heliocentric distance to M31.

\begin{figure*}
\centering
\includegraphics[scale=0.95]{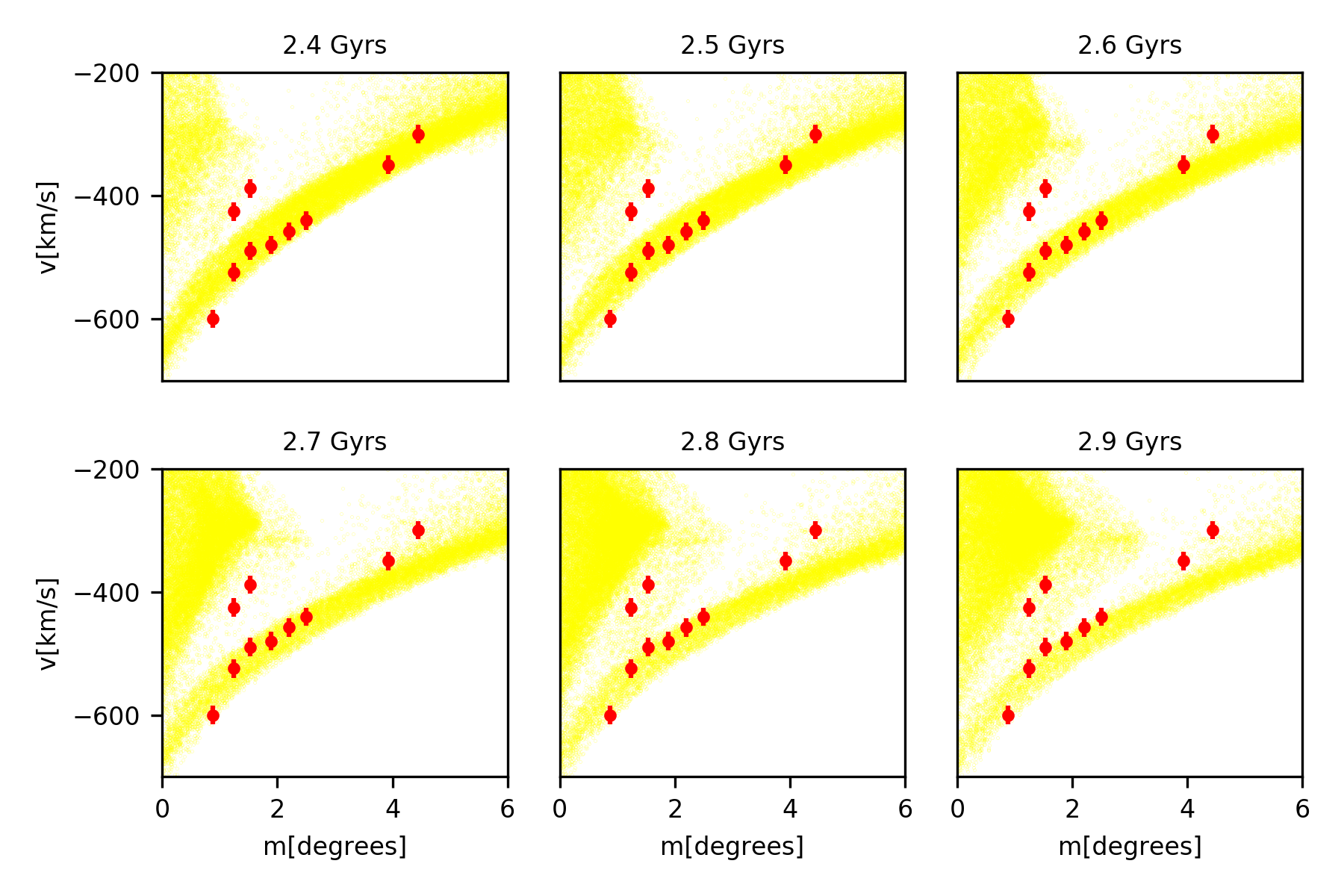}
\caption{Radial velocities along the GSS as a function of distance for the time interval between 2.4 and 2.9 Gyrs. Over-plotted thick red dots are the observed values from \citet{Ibata2004}, \citet{Guhathakurta} and \citet{Gilbert2009} .}
\end{figure*}


\begin{figure}
\centering
\includegraphics[scale=0.55]{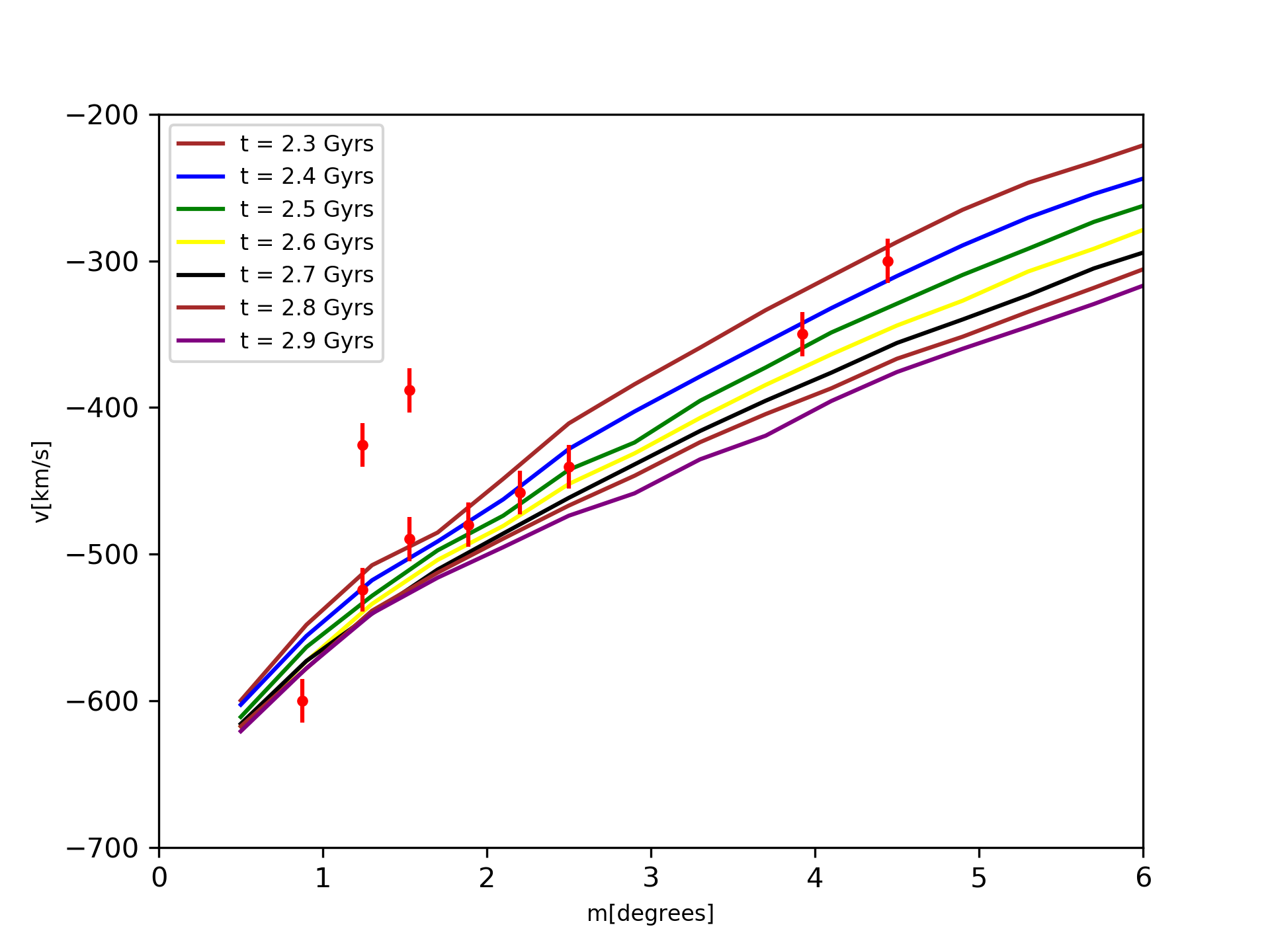}
\caption{The mean radial velocities along the GSS as a function of distance, for different times. Thick red dots are observed values from \citet{Ibata2004}, \citet{Guhathakurta} and \citet{Gilbert2009}.}
\end{figure}

\begin{figure}
\centering
\includegraphics[scale=0.55]{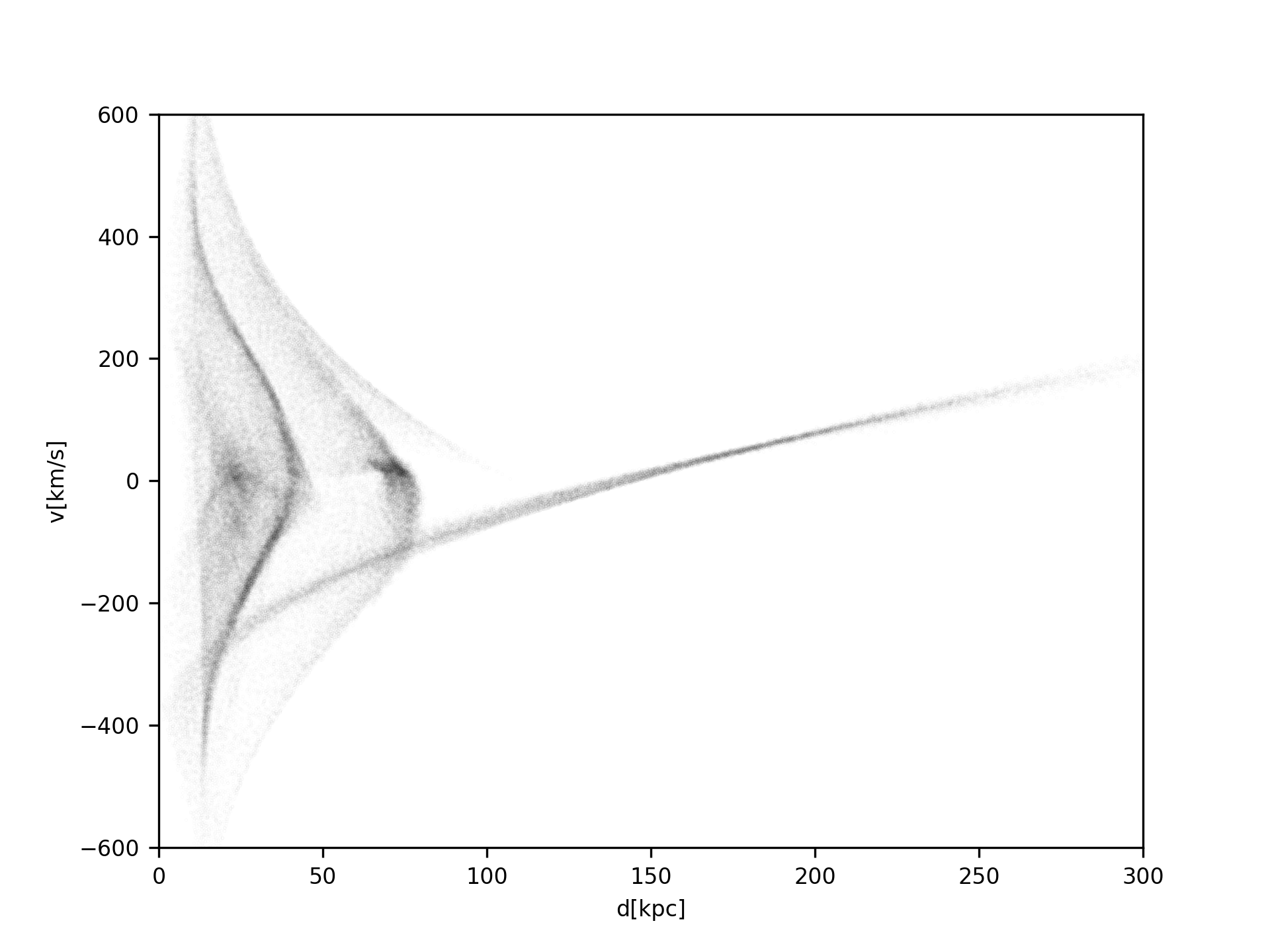}
\caption{The phase-space plot in d-$v_{rad}$ plane. We can see the GSS, W, and NE shelves. In the region of the NE shelf is the remnant of the progenitor. These results are in agreement with previous studies (\citealp{Fardal2013, Sadoun}).}
\end{figure}

\subsubsection{Distances and velocities along the Stream}
The depth and velocities of stars in the fields observed along the stream can be matched to the same properties of the particles in the simulated stream. We calculate heliocentric distances along the stream and compare them to the observed values in the fields given in \citealp{McConnachie, Conn}. Figure 2 shows distances of particles in our simulation from the center of M31 (depth). Over-plotted with bars are the observed values for the same distances in the observational fields of \citet{McConnachie} and \citet{Conn}. Position of the particles are translated in coordinates m and n, where m is a coordinate along the stream and n across the stream \citet{Fardal2007}:

\begin{equation}
m=0.504 \times \xi - 0.864 \times \eta \\
n=- 0.864 \times \xi - 0.504\times \eta
\end{equation}

\noindent Panels show the time snapshots into the merger when the GSS is already a formed structure. For the entire interval of 2.4 to 2.9 Gyrs, distances reproduced in our simulation are a good match to the observations.

This can also be seen in Figure 3 where we present the mean depth for different moments of the merger event. As already presented in Figure 3, we can see the match between observed and simulated values for all snapshots between 2.4 and 2.9 Gyrs, similarly to \citet{Sadoun}. As it is mentioned in previous works \citep{Sadoun}, the first observational field in \citet{McConnachie} is contaminated with stars from M31, and that is the reason for the lack of matching with simulated stream in that particular field.

\begin{figure}
\centering
\includegraphics[scale=0.55]{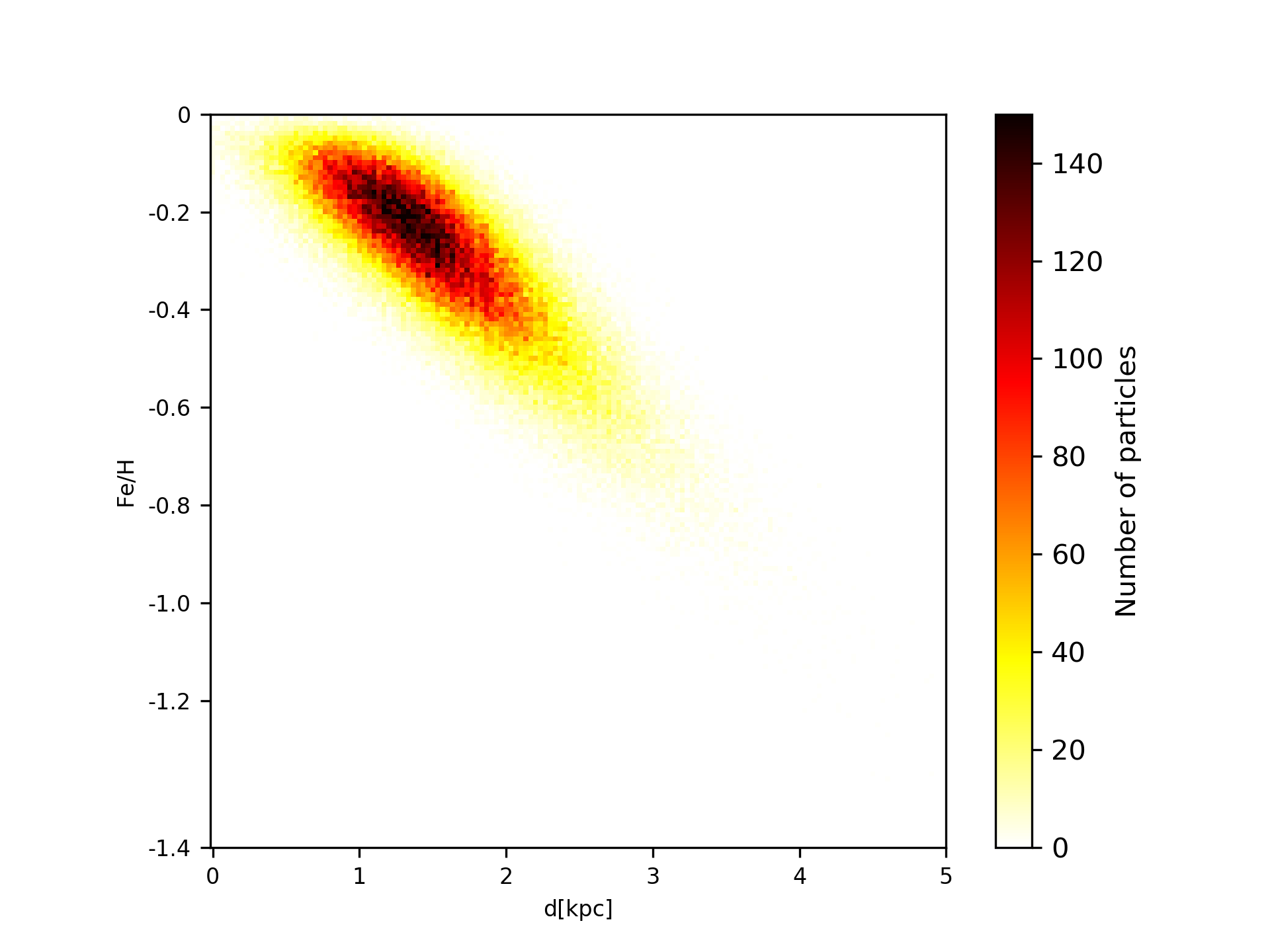}
\caption{Metallicity map for the GSS progenitor at the beginning of the simulation in the d - Fe/H plane, where d is the radial distance from the center of the dwarf galaxy.}
\end{figure}

Radial velocities along the GSS are presented in Figure 4. The observed values are from \citet{Ibata2004}, \citet{Guhathakurta}, \citet{Gilbert2009, Gilbert2018}. Velocity in the first observed field is overestimated, due to contamination from M31 stars in the observations, which can also be seen in the previous figures and in previous works \citep{Sadoun}. Panels show last 500 Myrs of GSS formation where we expect to find the best match between simulated and observed data. Thick red dots represent the observed values for the radial velocity and when compared to the radial velocities of the particles in our simulation, the best match is for snapshots at 2.4 and 2.5 Gyrs into the merger. This is in discrepancy with results from \citet{Sadoun} as their best match is for 2.7 Gyrs into the merger. However, one can notice that this difference occurs only in the last two observational fields beyond 4 degrees distance, while in the inner parts of the GSS, match is present at all times.

We plotted the mean radial velocity in Figure 5, confirming the results seen in Figure 4, showing even more clearly that the best match between observed and simulated data is moments at 2.4 and 2.5 Gyrs into the merger.

In the same merger event are formed NE and W shelves beside the GSS (\citealp{Fardal2007, Fardal2013, Sadoun}). These structures can be seen in phase-space, presented in Figure 6. In the same figure, we can see the remnant of the GSS in the region of the NE shelf. Our model well reproduces the shelves and the scenario of their formation is in agreement with previous studies (\citealp{Fardal2007, Fardal2013, Sadoun}). We also show particles in $d-v_{rad}$ plane for the best match of modeled metallicity with observed one, after 2.9 Gyr into the merger.   

\subsection{Results of Monte Carlo simulations}
Starting with the initial metallicity distribution in the GSS progenitor, and evolving it through the merger with M31, we have found a final metallicity distribution in the GSS. We compare this resulting distribution to the observed one \citet{Conn}. We also found additional constraint on the timescale of the merger event. The metallicity distribution together with morphological properties of the GSS give a timescale of 2.9 Gyrs, which is a slightly different timescale then suggested in \citet{Sadoun}, which is 2.7 Gyrs. 


\subsubsection{Metallicity distribution}
Several previous studies have focused upon the metallicities in the GSS from various observational programs. In \citet{Conn} and \citet{Cohen} we can see values for metallicity along the stream, given in 19 observed fields. There are  two gradients along the stream: in the inner part of the GSS metallicity increases from -0.7 to -0.2, and then decreases in the outer part, reaching the value of -1. With Monte Carlo simulations described in Section 2.4, we have tested if a linear function for the initial metallicity distribution in the dwarf galaxy, can explain two gradients in the metallicity along the GSS. Metallicity gradient $\Delta$FeH from the function 
FeH(r) = $\Delta$FeH $\times$ r is -0.3. We also tested different boundaries and different gradients from -0.1 to -0.5. These values are given in observations of \citet{Koleva2009a, Koleva2009b} and theoretical work of \citet{Kirihara2017}. The comparison of resulting metallicity distribution was made with results given in \citet{Conn}.

In Figure 7 we have represented the initial metallicity distribution, at t=0 Gyrs in the progenitor galaxy. As we can see from the colour density map, we assume that metallicity increases towards the central region of the galaxy.

In previous sections, we have shown that radial velocity and depth can be used to determine the formation time of the GSS. In a similar manner, the metallicity distribution along the stream at different times during the simulation can be matched to the observed one in order to predict the formation time of the GSS. Figure 8 shows simulated metallicity distribution (in red) in nine snapshots between 2.2 and 3.0 Gyrs. over-plotted with black dots is the observed metallicity in 19 fields of \citet{Conn}. The metallicity distribution evolves with time, because of the GSS dynamical evolution under the influence of dynamical friction and tidal effects. After 1000 Monte Carlo realisations of distance - metallicity parameter space, the resulting metallicity at each distance point along the stream, has a scatter around the mean metallicity value. This is represented by the thickens in the simulated metallicity distribution in Figure 8 (red lines).  

\begin{figure*}
\centering
\includegraphics[scale=0.75]{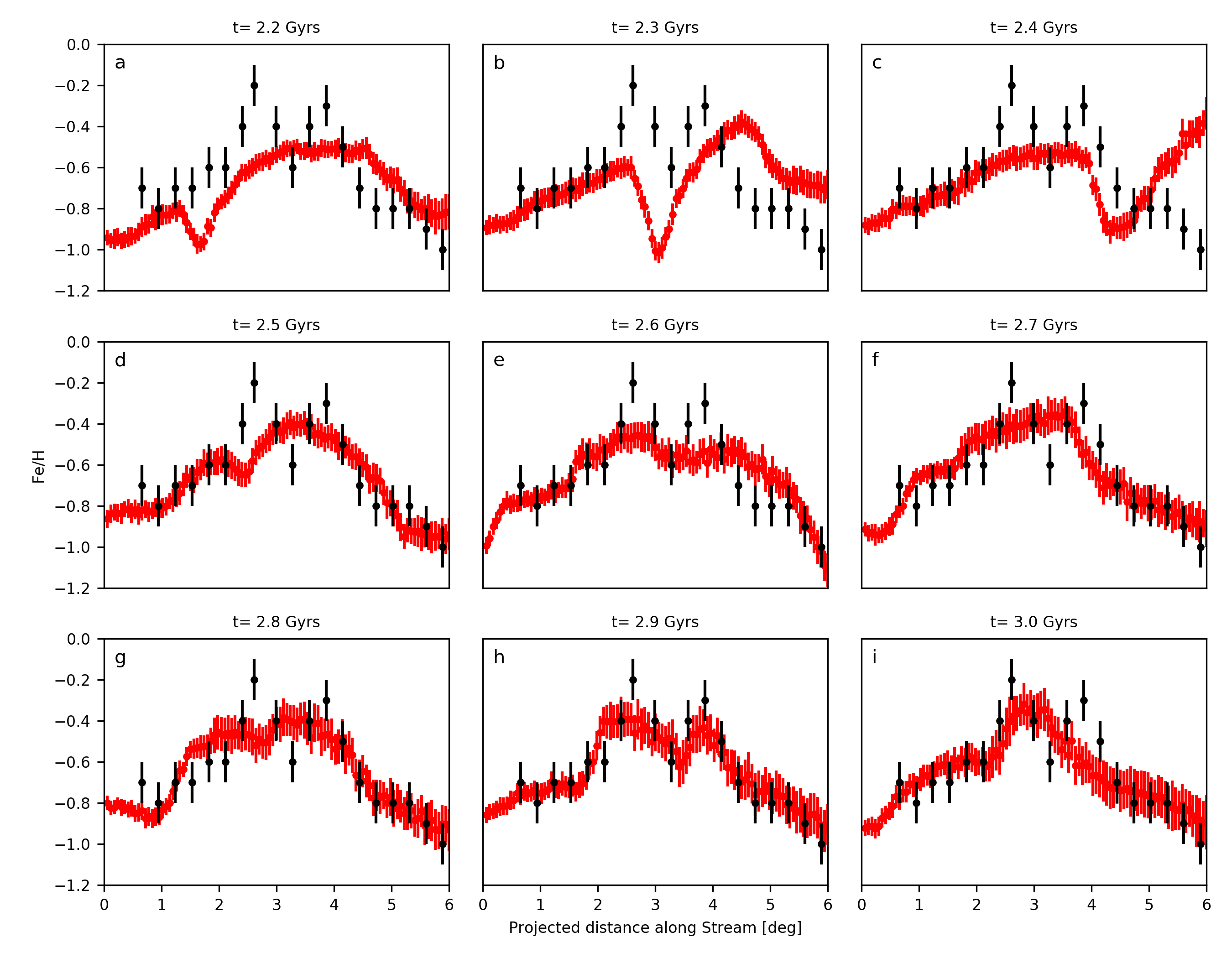}
\caption{Metallicity distribution along the stream from 2.2 Gyrs to 3.0 Gyrs. Red dots with error bars represent metallicity from Monte Carlo simulation and black dots are results given in \citet{Conn}. For a different time, we have different quality of matching.}
\end{figure*}

\begin{figure}
\centering
\includegraphics[scale=0.55]{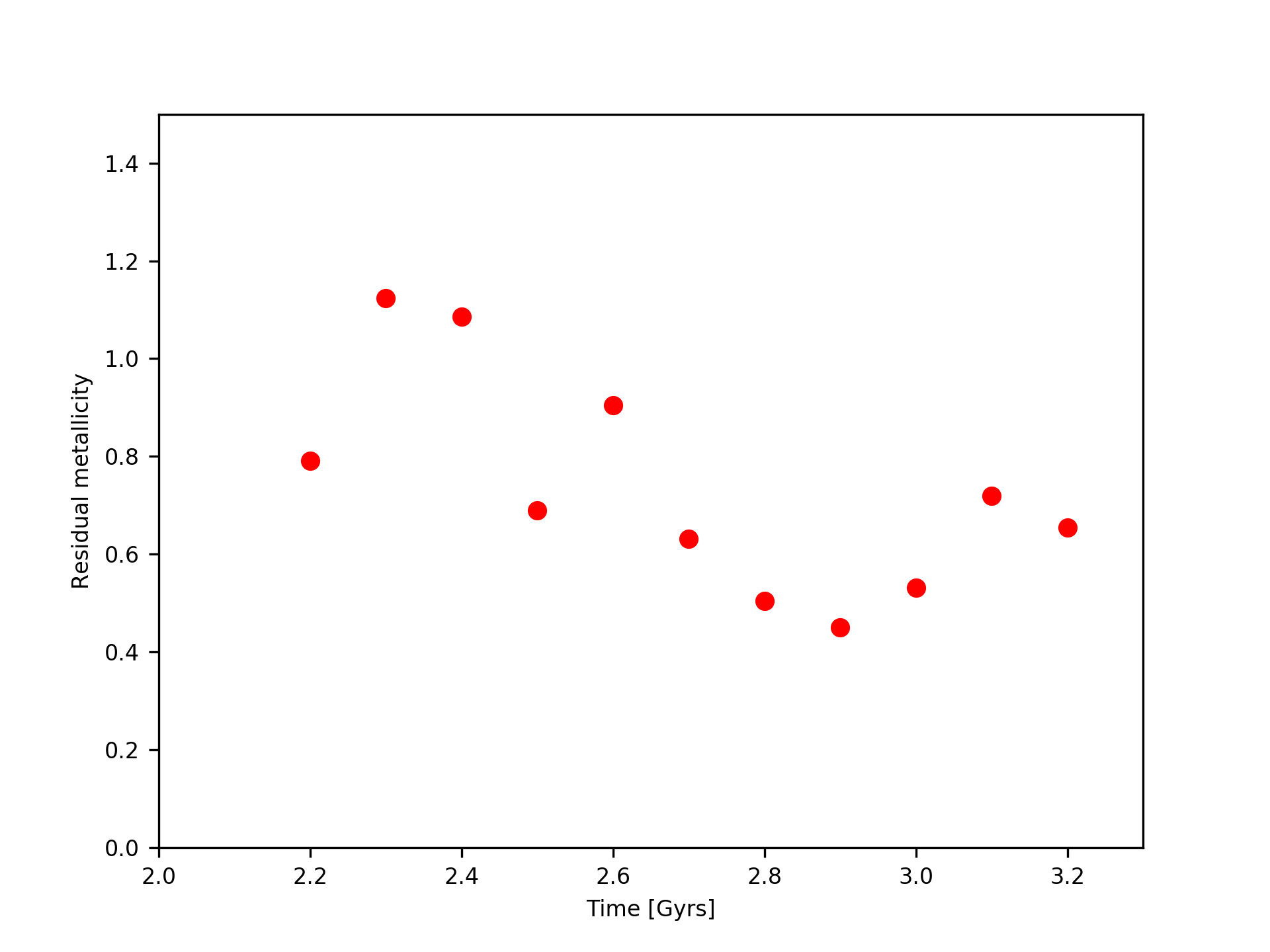}
\caption{Residual metallicity for different time intervals. Our model suggests that the best agreement is 2.9 Gyrs after merger event started.}
\end{figure}

Next we find the moment of the best match between simulations and observation by calculating the minimum squared difference between simulated and observed metallicities in Figure 8. 
\begin{equation}
\rm FeH_{residual} = \sum_{i=1}^{19} \sqrt{FeH_{MC}(i)^{2} - FeH_{observed}(i)^{2}}
\end{equation}
We ran our simulations slightly beyond 3.0 Gyrs and we calculated the residuals between 2.2 Gyrs, and 3.2 Gyrs. We can see in Figure 9 that the best agreement between simulated and observed metallicity distributions is at 2.9 Gyrs into the merger, since at that moment, residual in the metallicity has the smallest value. 

\begin{figure}
\centering
\includegraphics[scale=0.55]{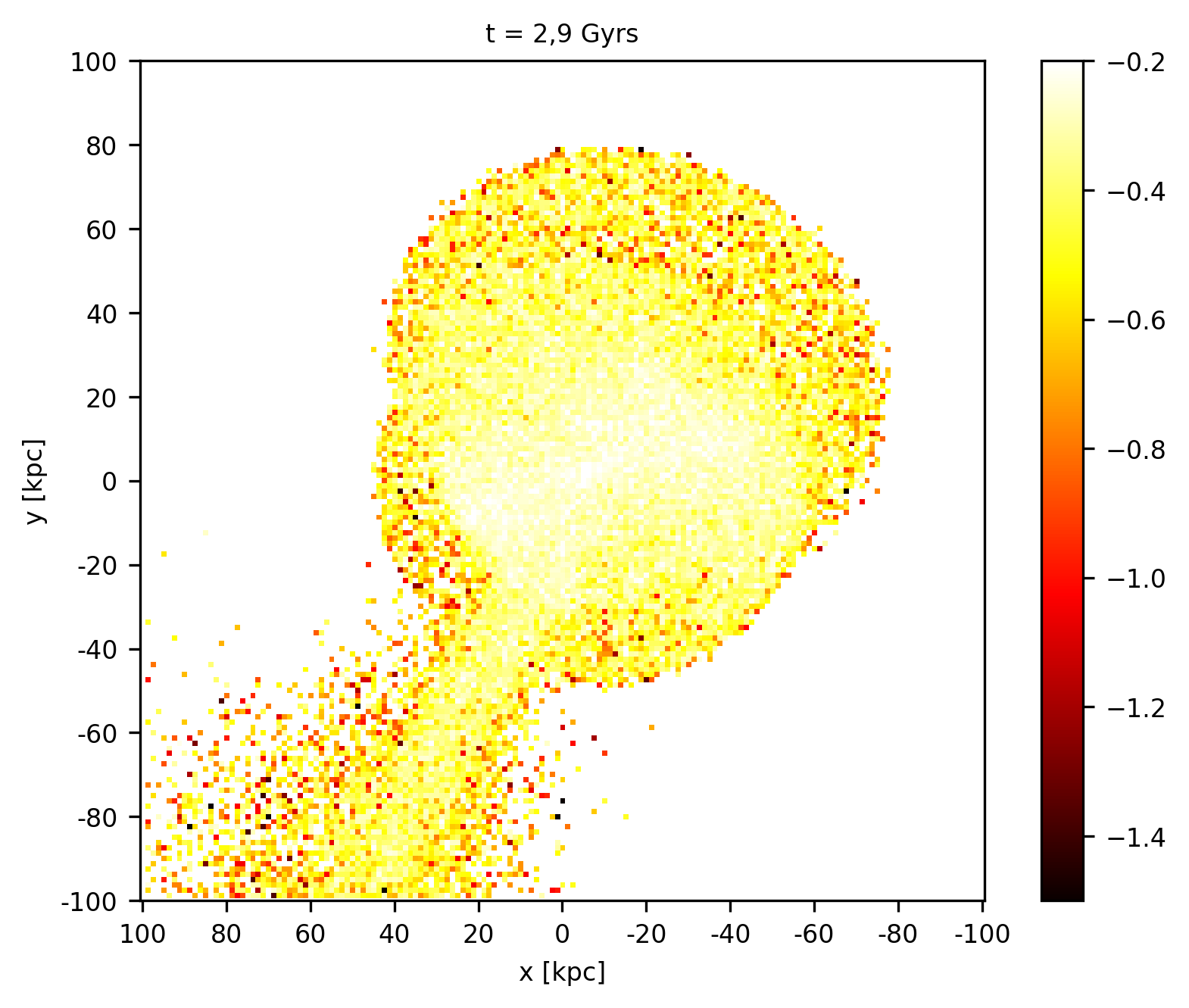}
\caption{Metallicity map of the GSS after 2.9 Gyrs. We can see lower metallicity vaules in the outer part of the GSS.}
\end{figure}

\begin{figure*}
\centering
\includegraphics[scale=0.75]{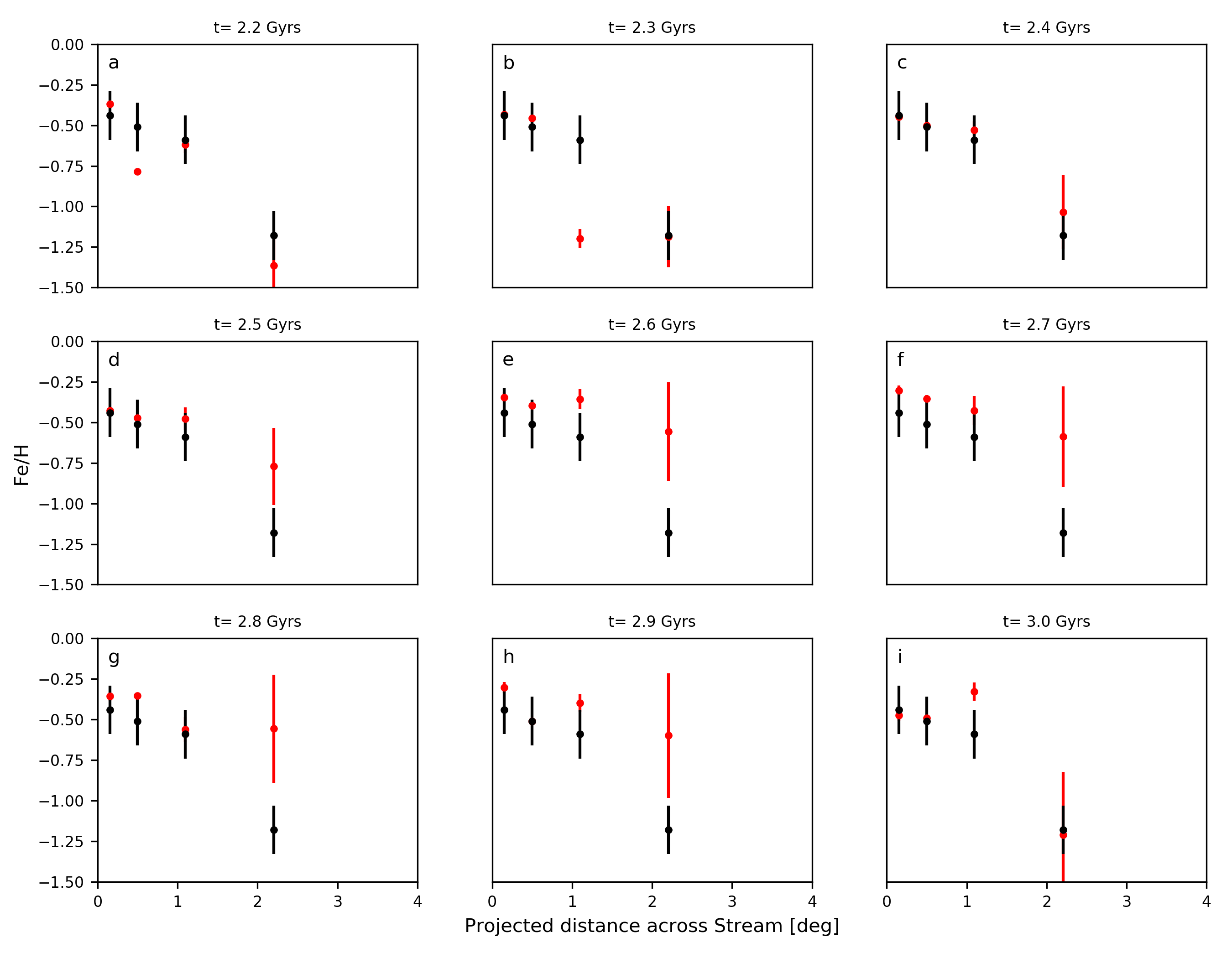}
\caption{Metallicity across the Stream. Red dots are simulated values, and black dots are the observed values from \citet{Gilbert2009}.}
\end{figure*}


Panel h in Figure 8 at 2.9 Gyrs represents the best match between simulated and observed metallicity. This can be seen for the most fields, but the largest disagreement is in field 8, for the largest value of the metallicity along the stream. There are different values for this field in previous works. \citet{Cohen} have reported a smaller value for this field than the one given in \citet{Conn} which would provide an even better match to our simulated metallicity. In our model, value of -0.2 is also out of predicted interval for that projected distance along the GSS.

If we look at the metallicity map for the best match metallicity at 2.9 Gyrs in Figure 10, we can see higher metallicity values in the core region of the GSS, and lower metallicity vaules in the outer part of the GSS. Observations show that besides the metallicity gradient along the stream, there is an also a gradient perpendicular to the stream (\citealp{Ibata2007, Gilbert2009}). Metallicity decreases perpendicularly from the center of the stream towards its outskirt. Figure 11 shows this behavior in our simulated stream.

\begin{figure*}
\centering
\includegraphics[scale=0.95]{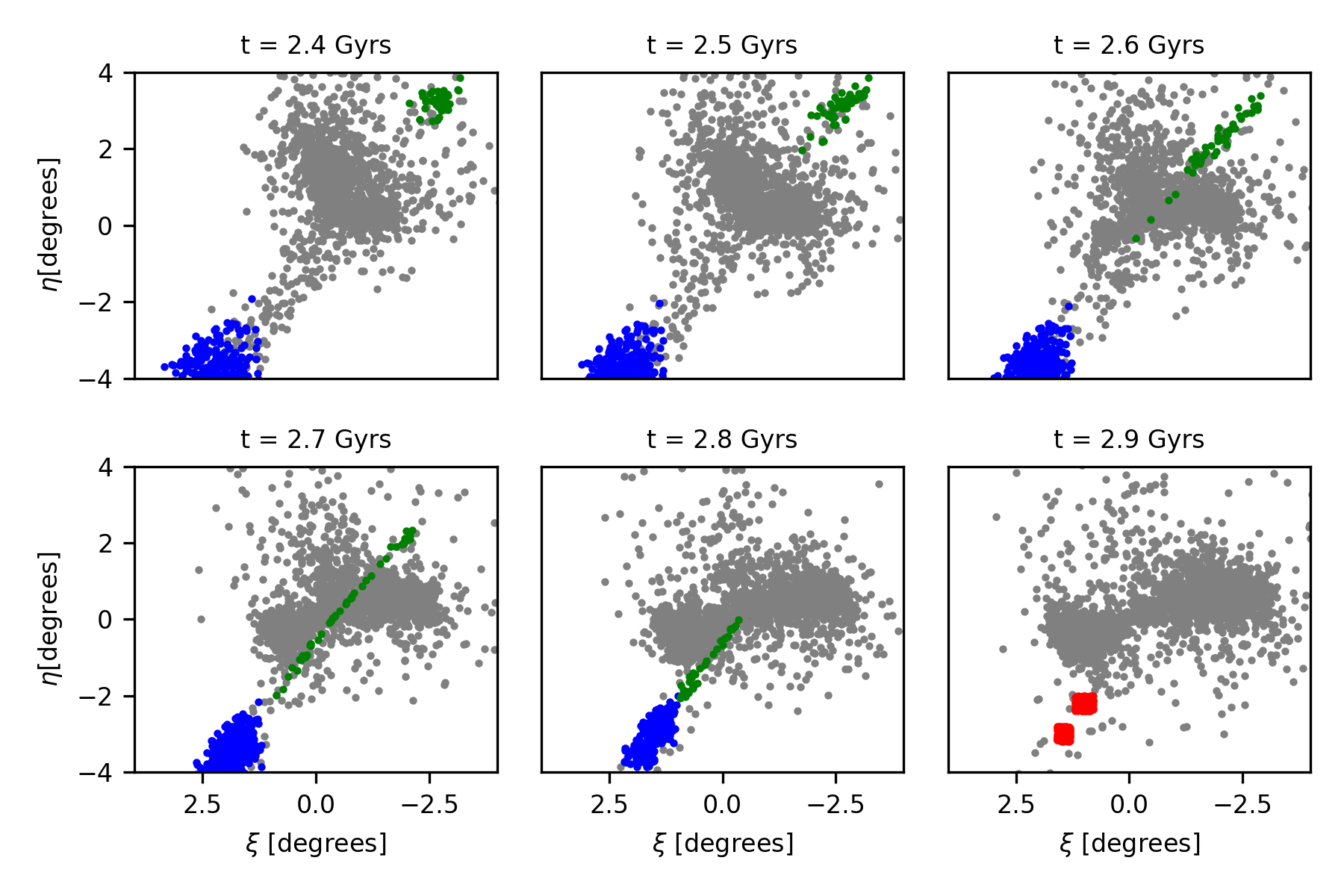}
\caption{Particle motion in opposite directions along the stream. Blue dots represent stars moving in NW direction, while green dots represent stars moving in SE direction. These two groups of particles are from the region where we can see peaks in metalicity distribution and they are represented with red dots in the last panel.}
\end{figure*} 


\subsubsection{Peaks in metallicity distribution}
We investigated the origin of the double peaked metallicity distribution and found that it originates from distinct kinematically groups.
Two kinematically distinct groups of stars are best represented by their coordinates in Figure 12. Blue points represent stars moving in North West (NW) direction, while green points represent stars moving in South East (SE) direction; metallicity peaks occur at chance as temporary structures when stars from these two groups cross their paths.
\begin{figure*}
\centering
\includegraphics[scale=0.8]{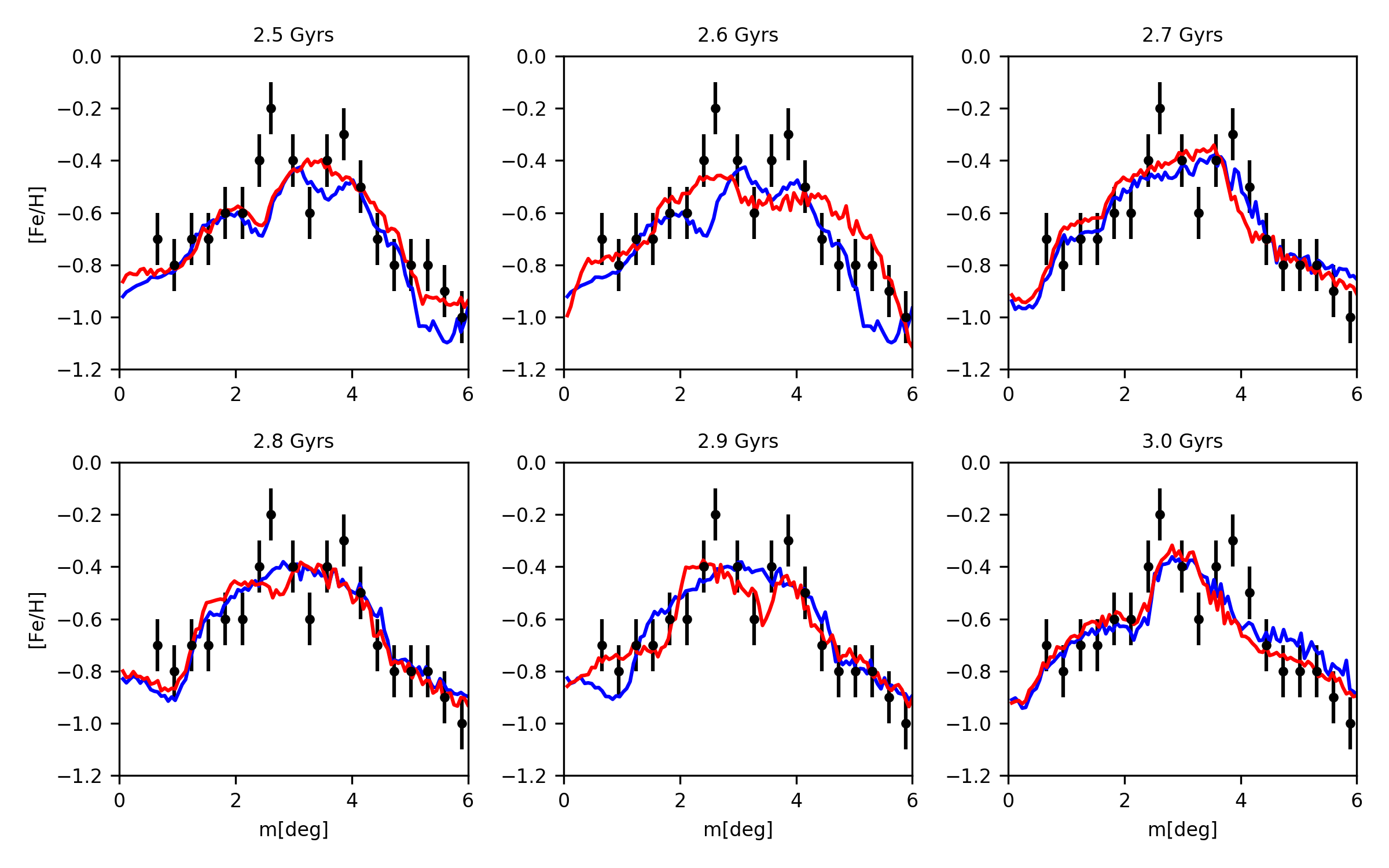}
\caption{Metallicity distribution along the stream for six snapshots. Observed metallicity distributions are represented by black dots, and simulated metallicity distributions in thick red line, corresponding to panels d, e, f, g, h, and i in Figure 8. With blue, thick line, metallicity distributions but only for particles moving in NW direction are presented.}
\end{figure*}

\begin{figure*}
\centering
\includegraphics[scale=0.9]{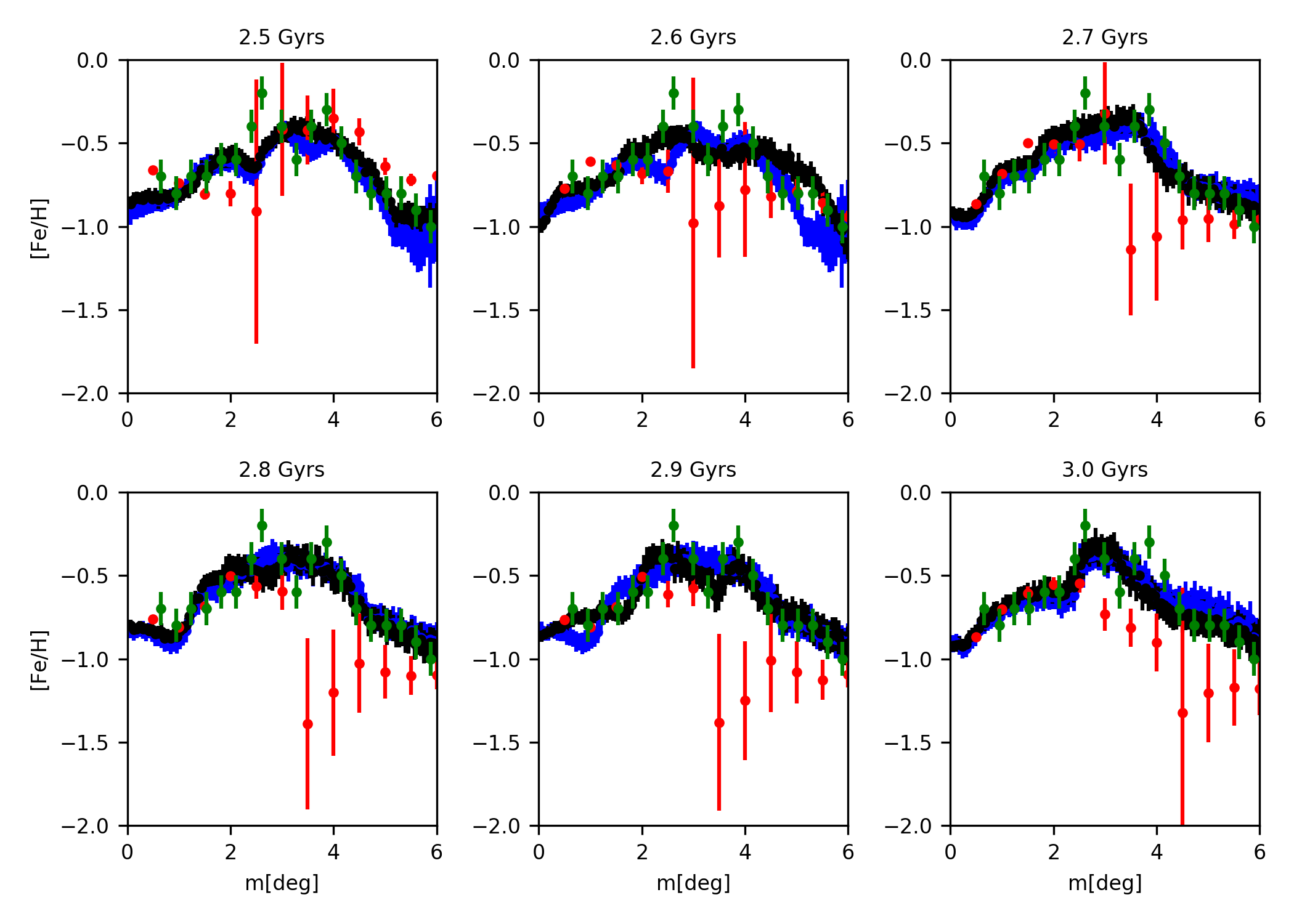}
\caption{Metallicity distribution along the stream for six snapshots. Observed metallicity distributions are presented by green dots, and simulated metallicity distributions in thick blue line, corresponding to panels d, e, f, g, h, and i in Figure 8, but only for particles moving in NW direction are presented. Red dots are for particles moving in SE direction, and with black thick line are presented all particles and their cumulative influence on final double-peaked distribution}
\end{figure*}

The observed metallicity distribution along the GSS shows two peaks, with no agreement on their origin 
(\citealp{Conn,Cohen}). It is also not clear if these peaks are permanent or transient structures. Since we can evolve metallicity along the stream, our simulations are the perfect tool to study the existence of metallicity peaks.  We find that the shape of metallicity distribution is guided by the dynamics of the GSS formation. The GSS forms through the shredding of the material from the dwarf galaxy merging with M31. Due to the merger circumstances, dwarf galaxy passes through the center of M31 on almost radial orbit only to be pulled back shortly. This repeats more than once. Each time the dwarf passes through the center it looses significant amount of stars. The consequence is that although all detached stars are moving along the stream, they are not all moving in the same direction. 

In Figure 12 we can see that some particles are moving in SE direction away from the center, while others are moving in NW direction, toward the center. This particles have different velocities in the $\xi$ - $\eta$ plane, but also different line-of-sight velocities. Particles that are moving in SE direction have positive lign-of-sight velocities $\sim$ 30 km/s and, mean proper motion $\sim$ 0.1 mas/yr, while particles that are moving in oposite, NW dorection, have negative lign-of-sight velocities $\sim$ -420 km/s and, mean proper motion $\sim$ 0.02 mas/yr. Figure 13 shows observed metallicity distributions in black dots, and simulated metallicity distributions in thick red line, for all particles, corresponding to panels d, e, f, g, h, and i in Figure 8. With blue, thick line, it also shows metallicity distributions but only for particles moving in NW direction. We can see two peaks only when all particles are included, so inner dynamics of the stream, precisely the fact that we have at least two kinematic groups of particles, is responsible for this kind of distribution with two peaks. 

The comparison of metallicity distributins when all particles are included and when we include only groups that are moving in NW or SE directions is represented in Figure 14. The dynamical evolution of the stream, in other words, opposite directions of particle motion, makes peaks a temporary structures, but general distribution with maximum metallicity in the central region of the GSS is still present.

\section{Discussion and Conclusions}
In this paper we have investigated the metallicity distribution in the Giant Stellar Stream in the M31 galaxy. The formation of the stream is still an open question and we considered the possible scenario of a single merger event. Different morphologies of the dwarf galaxy were tested in previous works, and in this work a dSph was considered. 
We ran pure N-body simulation due to the absence of gas to confirm the dynamical history of the merger event by comparing with observational criteria including the distance from the center of M31, velocity distribution, and spatial orientation of the stream. As it was found in previous works, we can also confirm that the timescale of the merger event is between 2 and 3 Gyrs.

After the first  observations along Stream were done by (\citealp{Conn, Cohen}) we were able to constrain distribution in progenitor and compare its initial distribution with the final one in the Stream. We reproduced distribution presented in \citet{Conn}, as it is formed after the merger of the satellite galaxy and M31. We took a linearly decreasing function for metallicity gradient in the dwarf galaxy at the beginning of the simulation, and with Monte Carlo simulations we attached metallicity values for particles and reproduced observed distribution for the time interval between 2 and 3 Gyrs after the beginning of the simulation. The best agreement between simulated and observed metallicity distribution we got for 2.9 Gyrs after the beginning of the merger. In \citet{Ibata2007} and \citet{Gilbert2009} core to envelope metallicity is presented. In this work, we were also able to reproduce core-envelope metallicity distribution and showed more metal-rich core. For the core-envelope gradient we found the best match with observed data for 2.4 Gyrs (Figure 11). In direction across the Stream, we compare our results for only four observed values, unlike metallicity distribution along the Stream. Nevertheless, we reproduced observed metallicity distributions in both directions, along,  and across GSS, for the formation time between 2 and 3 Gyrs.

We showed that dynamic properties could be responsible for two observed peaks of metallicity in the central region of the GSS. We traced particles from those peaks and deduced that the inner dynamics of the stream are influenced by at least two groups of stars: one that moves toward the central part of the GSS in NW direction, and the other that moves in opposite SE direction. This leads to the conclusion that two peaks are temporary structures in metallicity distribution, but general distribution along the stream with maximum metallicity in the central part and lower at ends is present on the larger timescale. In \citet{Kalirai2006a} two kinematically cold components are represented. We cannot connect our results in double-peaked metallicity distribution with second cold kinematically component, as these two components were observed in H13s field (\citealp{Kalirai2006a, Cohen}) on $R_{proj}$ = 20 kpc which corresponds to field in \citet{Conn} at $m = 1.53$ degrees, and double-peaked in metallicity is between 2.5 and 4 degrees (\citealp{Conn, Cohen}). Nevertheless, observations of second cold kinematically component in the field H13s and double-peaked metallicity in observations along Stream unveil a very complex kinematical picture.    

Finally, we found that the initial metallicity distribution in progenitor can be responsible for the present metallicity distribution in the GSS. This could be an additional constrain for the dynamical properties of the merger event. Our model also predicts the NE and W shelves as structures formed in the same merger event as GSS. Metallicity distribution of shelves will be subject of the following paper.
\section*{ ACKNOWLEDGEMENTS} 
This work was supported by the Ministry of Education, Science and Technological Development of the Republic of Serbia through contracts: 451-03-9/2021-14/200104 and 451-03-9/2021-14/200002.

\section*{Data Availability}
Data used in the paper will be made available on a reasonable request to the author. 



\bibliographystyle{mnras}
\bibliography{GSS} 




\label{lastpage}
\end{document}